\documentclass[twocolumn,showpacs,preprintnumbers,amsmath,amssymb,superscriptaddress]{revtex4}

\usepackage{calc}
\usepackage{float}
\usepackage{graphicx}
\usepackage{appendix}
\usepackage{bm}
\def \equi#1{\mathrel{\mathop{\kern 0pt\sim}\limits_{#1}}} 
\scrollmode

\begin{document}

\title{Fluctuations and correlations
of a driven tracer  in a hard-core lattice gas}

\author{O. B\'enichou}
\affiliation{Laboratoire de Physique Th\'eorique de la Mati\`ere Condens\'ee, CNRS UMR 7600, case courrier 121, Universit\'e Paris 6, 4 Place Jussieu, 75255
Paris Cedex}

\author{P. Illien}
\affiliation{Laboratoire de Physique Th\'eorique de la Mati\`ere Condens\'ee, CNRS UMR 7600, case courrier 121, Universit\'e Paris 6, 4 Place Jussieu, 75255
Paris Cedex} 

\author{G. Oshanin}
\affiliation{Laboratoire de Physique Th\'eorique de la Mati\`ere Condens\'ee, CNRS UMR 7600, case courrier 121, Universit\'e Paris 6, 4 Place Jussieu, 75255
Paris Cedex}

\author{R. Voituriez}
\affiliation{Laboratoire de Physique Th\'eorique de la Mati\`ere Condens\'ee, CNRS UMR 7600, case courrier 121, Universit\'e Paris 6, 4 Place Jussieu, 75255
Paris Cedex}

\date{\today}

\begin{abstract}
We consider  a driven tracer particle (TP) in a bath of hard-core particles undergoing continuous exchanges
with a reservoir. 
We develop an analytical framework which allows us to go beyond 
the standard force-velocity relation used for this minimal model of active microrheology 
   and quantitatively analyze, for {\it any} density of the bath particles, the fluctuations of the TP position 
   and their correlations with the occupation number of the bath particles. 
   We obtain an exact Einstein-type relation which links 
   these fluctuations in the absence of a driving force and
    the bath particles density profiles  
    in the linear  driving regime. 
    For the one-dimensional case we also provide an approximate but very 
    accurate explicit expression for the variance of the TP position 
    and  show that it can be a non-monotoneous function of the bath particles density: 
    counter-intuitively, an increase of 
    the density may increase the 
   dispersion of the TP position. 
   We show that this non trivial behavior,  which could in principle be observed in active microrheology experiments, is induced by subtle  cross-correlations quantified by our approach.

  \end{abstract}


\maketitle

\section{Introduction}

Active probe-based microrheology, which
monitors the position of a tracer particle (TP) driven by an external force, 
has become a powerful experimental tool for the analysis 
of different systems in physics, chemistry and  biology  
(see, e.g., Ref. \cite{Wilson:2011a}). 
In the constant-forcing microrheology  
setup, a microscopic particle embedded in a 
sample is actively manipulated by applying a known force. One  
measures then the response of the TP and potentially the 
microstructural deformations 
of the medium 
to learn about its microrheological properties. 
From the theoretical point of view, an unresolved issue, which is disregarded  in the available continuous analytical approaches, is to take into account explicitly  the discreteness or "granularity" of the medium \cite{Wilson:2011a}.  This aspect  becomes crucial when the probe and the medium particles have comparable sizes, e.g. in colloidal suspensions. In this regime,  both fluctuations of the probe
  that can not be described correctly if the medium is treated as a continuous bath and super-diffusion regimes have been reported \cite{Wilson:2011,Winter:2012,2012arXiv1210.1081S,2012arXiv1211.0674B,Harrer:2012}

Within a broader context, modeling the response of a medium to a perturbation created by
a tracer particle, biased by an external force,  
is a ubiquitous problem in physics, which has 
been the subject of a large number of theoretical works \cite{Marconi:2008,Cugliandolo:2011,vulp}. The resulting stochastic dynamics of the whole system is however a many-body problem which is difficult (or even impossible) to solve even in the simplest case when the particle-particle interactions are a mere hard-core. For that reason, in most approaches the microscopic structure of the
bath is not taken into account
explicitly, and the response  functions are determined instead
by using some effective bath dynamics, modelled via  Langevin or generalized Langevin \cite{Lizana:2010} equations (see also Refs. \cite{Marconi:2008,Cugliandolo:2011,vulp}). While these approaches are rather 
efficient, they cannot 
account for the detailed correlations between the 
tracer particle and the fluctuating density profiles of the bath particles.  In this paper, we show that these
 cross correlations are of crucial importance and can lead to  non trivial effects  beyond the usual analysis of the force-velocity relation. We predict in particular  a non monotonic behavior of the variance of the tracer position with  the density of the bath, that in principle could be observed in active microrheology experiments.

Our analysis relies on a model of driven tracer diffusion in a hard-core lattice gas, which appears as  a minimal  model of active microrheology that explicitly takes into account  the dynamics of a bath of discrete particles:  A TP driven by an external force performs a  biased lattice random walk in a bath of hard-core particles, which themselves perform symmetric random walks constrained by the condition of a single occupancy of each lattice site. 


At the theoretical level, the one dimensional version of this model is related to several well known models of out of equilibrium statistical physics. For example, in the absence of  external driving, it reduces to the single file diffusion problem  \cite{Harris:1965}. In the absence of external driving and when additional adsorption/desorption processes with a reservoir of particles are considered, it corresponds to the so-called dynamical percolation \cite{Druger:1983a,Benichou:2000la}. Last, in the case of a constant external forcing experienced by {\it all} the particles, it identifies with the asymmetric exclusion process, which has now become a paradigmatic model, both in the absence (see \cite{Chou:2011} for a recent review)  or in the presence \cite{Parmeggiani:2003} of adsorption/desorption processes.
This model of driven tracer diffusion  has been investigated both in the physical \cite{Burlatsky:1992a,Burlatsky:1996b,Benichou:1999c,Benichou:2000gd,Benichou:2001a} and in the mathematical \cite{Landim:1998a,Komorowski:2005a} literatures. Most of the obtained results, including proofs of the Einstein relation,  are limited to the large time behavior of 
the mean position   $\langle X_{tr}\rangle$ of the tracer particle  
and the stationary density profiles of the bath $\langle\eta_{\bf r}\rangle$, where $\eta_{\bf r}$ stands for the occupation number of the site ${\bf r}$  of the lattice, equal to 1 if occupied by a bath particle and 0 otherwise. 
A general method to analyze the   fluctuations of the tracer position $\langle \delta X_{tr}^2\rangle$ and  the cross-correlation functions of the form 
$\langle \delta X_{tr} \delta \eta_{\bf r}\rangle$ is still lacking \footnote{Note that only  $\langle \delta X_{tr}^2\rangle$ has been considered in \cite{2012arXiv1211.0674B}, and the analysis was limited to  the specific regime of high density of bath particles}.

Here, for this minimal model of active microrheology we develop a  theoretical framework allowing to analytically determine these fluctuations  in all regimes of bath particles density, and in principle for arbitrary dimension.  More precisely, our main results obtained within this formalism are (i) An exact Einstein-type relation satisfied by the cross-correlations in the absence of a driving force and the density profiles of the bath particles in the linear-response  regime.  (ii)  An explicit  approximate but accurate expression for the variance of the tracer position  in a one-dimensional case valid for any density. (iii) A finding that the variance of the position of the tracer can be a non-monotonic function of the density of bath particles, 
meaning that, counter-intuitively, an increase of 
the density of hard-core particles can increase the dispersion of the tracer position. We show that,  in fact,  subtle correlations  between the tracer and the bath particles 
are responsible for this intriguing behavior.  We anticipate that this striking effect could in principle be observed experimentally in the context of active microrheology.

\section{The model}

\begin{figure}[h!]
\begin{center}
      \includegraphics[width=8cm]{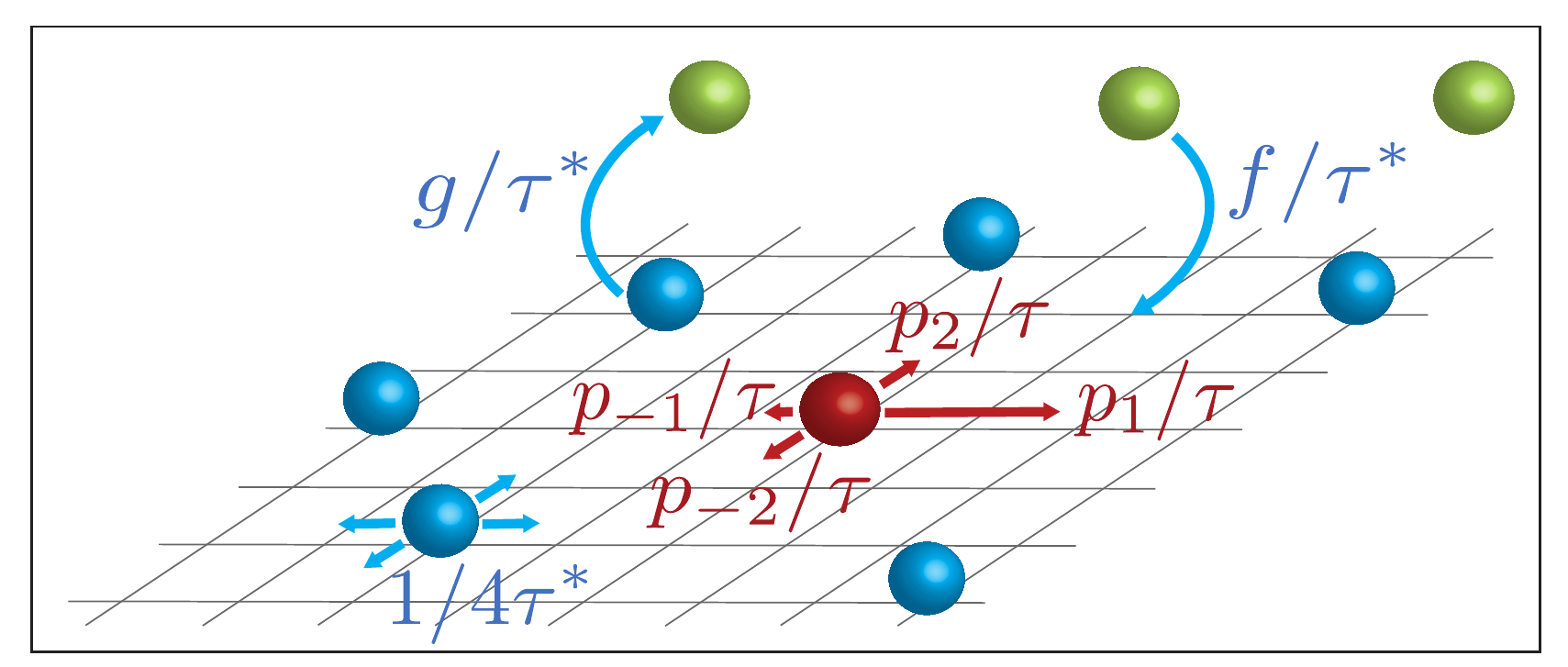}
\end{center}
\caption{Model notations in the two-dimensional (2D) case.}
\label{modele}
\end{figure}

Consider a $d$-dimensional  
hypercubic lattice of spacing $\sigma$
in
contact with a reservoir of particles kept at a constant chemical potential (see figure \ref{modele}).  
Suppose next that the particles in the reservoir 
may adsorb 
onto vacant lattice sites at a fixed rate $f/\tau^*$. 
The adsorbed particles may move randomly along the lattice by  
hopping at a rate $1/ 2d \tau^*$ to any of $2d$
neighboring lattice sites,
which process is 
constrained by a hard-core exclusion preventing multiple occupancy of any of the sites.
The adsorbed particles may 
 desorb from the lattice  back to the reservoir
at rate $g/\tau^*$.
The occupancy of lattice sites is described by the time-dependent Boolean
variable $\eta_{{\bf r}}$, which assumes two values,  $1$, if the site ${\bf r}$ is occupied by an adsorbed particle, and $0$, otherwise.
Note  that the mean density
of the bath particles, $ \langle\eta_{\bf r}\rangle$, approaches as
$t\to\infty$ a constant value $\rho_s=f/(f+g)$ but
the number of particles on the lattice is not explicitly 
conserved in such a dynamics.

At $t = 0$ we introduce a TP, whose 
position at time $t$ is denoted as ${\bf R}_{tr}(t)$. The TP dynamics is different from that of the adsorbed particles in two aspects:
first, it can not desorb from the lattice and second,  
it is subject to
an external driving force  {\bf E}, which favors its jumps along the direction 
corresponding to the unit vector ${\bf e_1}$ of the lattice. Physically, such a  situation is realized in the context of active microrheology where the force on the TP is classically exerted by  magnetic tweezers \cite{Wilson:2011a}. 

The TP dynamics is defined in the usual fashion:
We suppose
that the tracer, which occupies the site ${\bf R}_{tr}$ at time $t$,
 waits an exponentially distributed time with mean
$\tau$, 
and then attempts to hop onto one of $2d$ neighboring sites, ${\bf
R}_{tr} + {\bf e}_{\boldsymbol \nu}$, where ${\bf e}_{\boldsymbol \nu}$
are $2d$  unit 
vectors of the hypercubic lattice. 
The jump direction is chosen 
according to the probability $p_\nu$, which obeys:
\begin{equation}
p_\nu=\exp\left[\frac{\beta}{2}({\bf E \cdot e}_{
\nu})\right]\left/\sum_{\mu}\exp\left[\frac{\beta}{2}({\bf E \cdot e}_{ \mu})\right]\right.,
\label{defp}
\end{equation}    
where $\beta$ is the inverse temperature and $\sum_{\mu}$ denotes summation over all possible
orientations
of the vector ${\bf e}_{\mu}$. 
The hop is instantaneously fulfilled 
if 
the target site
is vacant at this moment of time; otherwise, i.e., if the target site
is occupied by any adsorbed particle, the jump is rejected and the
tracer remains at its position.

\section{Evolution equations}

We focus on the projection of the TP position  $X_{tr}\equiv {\bf R}_{tr}\cdot {\bf e}_1$, which  evolves in an infinitesimal time interval $\mathrm{d}t$ according to
\begin{equation}
\label{increment}
X_{tr}(t+\mathrm{d}t) = \left\{\begin{array}{ll}
X_{tr}(t)+\sigma,     \mbox{ w.p.  $\pi_1=p_1 (1-\eta_1)\mathrm{d}t$} \nonumber\\
X_{tr}(t)-\sigma,     \mbox{ w.p. $\pi_{-1}=p_{-1} (1-\eta_{-1})\mathrm{d}t$} \nonumber\\
X_{tr}(t),     \mbox{w.p. $1-\pi_1-\pi_{-1}$}
\end{array}
\right.
\end{equation} 
where w.p. stands for with probability. Averaging this equation, it is easy to find that the mean $\langle X_{tr}\rangle$ and the variance $\langle \delta X_{tr} ^2\rangle$ of the TP position verify:
\begin{equation}
\frac{\mathrm{d}}{\mathrm{d}t} \langle X_{tr}(t)\rangle=
\frac{\sigma}{\tau}\{p_1  (1-k_1)-p_{-1} (1-k_{-1})\},
\label{Xmean}
\end{equation}
and
\begin{subequations}
\begin{align}
\label{1}
&\frac{\mathrm{d}}{\mathrm{d}t}\langle \delta X_{tr}(t) ^2\rangle  =
\frac{\sigma^2}{\tau}(1-\rho_s)\\
\label{2}
&-\frac{\sigma^2}{\tau}\{p_1 (k_1-\rho_s)+p_{-1} (k_{-1}-\rho_s)\}\\
\label{3}
&-\frac{2 \sigma}{\tau}\{p_1  \widetilde{g}_1-p_{-1} \widetilde{g}_{-1}\}
\end{align}
\label{dispersion}
\end{subequations}
where
\begin{equation}
k_{\boldsymbol \lambda} \equiv\langle \eta_{\bf R_{tr}+{\boldsymbol \lambda}}\rangle \;{\rm and}\;
\widetilde{g}_{{\boldsymbol \lambda}}\equiv\langle\delta X_{tr}\delta\eta_{{\bf R}_{tr}+{\boldsymbol \lambda}}\rangle
\label{defgtilde}
\end{equation}
and 
\begin{equation}
k_{\pm1}\equiv k_{\pm {\bf e}_1}\;\;{\rm and}\;\;\widetilde{g}_{\pm1}\equiv \widetilde{g}_{\pm {\bf e}_1}
\end{equation}
Note that  the right-hand side of (\ref{1}) is a trivial mean-field term in which the effective jump frequency is given by $(1-\rho_s)/\tau$. In turn, the two other contributions are nontrivial: the first one (\ref{2}) originates from the  difference between the actual density of bath particles in the TP neighborhood and their unperturbed value $\rho_s$. The second one (\ref{3}) involves the correlations between the TP position and the occupation numbers in front and behind the tracer. As we proceed to show, these two non trivial contributions, referred to in the following as the density and the cross-correlation contributions, are responsible for  a non trivial behavior of the variance $\langle \delta X_{tr} ^2\rangle$.
 
Now, in order to calculate  $\langle X_{tr}\rangle$ and  $\langle \delta X_{tr} ^2\rangle$, we 
have to determine the density profiles 
 $k_{\pm1}$  and the cross-correlation functions $\widetilde{g}_{\pm1}$ at the  sites adjacent to the TP, which requires, in turn, the evaluation of 
the density profiles $k_{{\boldsymbol \lambda}}$ and the cross-correlation functions $\widetilde{g}_{{\boldsymbol \lambda}}$  for arbitrary $\boldsymbol \lambda$.
The evolution equations of these latter quantities can be deduced
 from the master equation (see appendix \ref{appendixA})   
but these equations are not closed. Actually, they  are coupled to the higher order
correlations functions, so that  one faces
the problem of solving an infinite hierarchy of coupled equations.  
Here we resort to the simplest non-trivial 
closure of the hierarchy, by using the following decoupling approximation:
\begin{eqnarray}
\label{prof}
&&\langle \eta_{\boldsymbol \lambda} \eta_{{\bf e}_{ \mu}} \rangle \approx \langle \eta_{\boldsymbol \lambda}\rangle \langle \eta_{{\bf e}_{ \mu}}\rangle\\
&&\langle \delta X_{tr}\eta_{ \boldsymbol\lambda} \eta_{{\bf e}_{ \mu}} \rangle\approx \langle \eta_{ \boldsymbol\lambda}\rangle \langle \delta X_{tr}\delta\eta_{{\bf e}_{ \mu}}\rangle + 
 \langle \eta_{{\bf e}_{ \mu}}\rangle\langle \delta X_{tr}\delta\eta_{ \boldsymbol\lambda}\rangle 
 \label{corr}
\end{eqnarray}
for ${\boldsymbol \lambda}\neq\{{\bf 0},\pm{\bf e}_1,\pm{\bf e}_{2}\ldots,\pm{\bf e}_d\}$. 
The approximation in Eq. (\ref{prof}) has already been used in Refs. \cite{Benichou:1999c,Benichou:2000gd,Benichou:2001a} and was shown to result in very good quantitative estimates of the TP mean position, while the one in (\ref{corr}) is new and, as we proceed to show, 
gives approximate expressions for the variance of the TP position which are 
in excellent agreement with numerical simulations. Note that these relations are natural mean-field type approximations, since they can be regarded as expansions at the leading order in the fluctuation parameter $\delta \eta_{\boldsymbol \lambda} \equiv \eta_{\boldsymbol \lambda} - \langle \eta_{\boldsymbol \lambda} \rangle$.

Relying on these approximations, we finally obtain  closed equations for $k_{{\boldsymbol \lambda}}$ and  $\widetilde{g}_{{\boldsymbol \lambda}}$ which hold for all ${\boldsymbol \lambda}$, except for  ${\boldsymbol \lambda}=\{{\bf 0},\pm{\bf e}_1,\pm{\bf e}_{2}\ldots,\pm{\bf e}_d\}$:
\begin{equation}
2d\tau^*\partial_tk_{{\boldsymbol \lambda}}=\tilde{L}k_{{\boldsymbol \lambda}}+2df, 
\label{systemek1}
\end{equation}and 
\begin{eqnarray}
2d\tau^*\partial_t\widetilde{g}_{{\boldsymbol \lambda}}&=&\widetilde{L}\widetilde{g}_{{\boldsymbol \lambda}}+\frac{2d\tau^*}{\tau}\sigma\left\{
p_1(1-k_{{\bf e}_1})k_{{\boldsymbol \lambda}+{{\bf e}_1}}\right.\nonumber\\
&-&\left.p_{-1}(1-k_{-{\bf e}_1})k_{{\boldsymbol \lambda}-{{\bf e}_1}}\right\}-\frac{2d\tau^*}{\tau}\sum_\mu p_\mu\widetilde{g}_{{\bf e}_\mu}\nabla_\mu k_{{\boldsymbol \lambda}}\nonumber\\
&-&\frac{2d\tau^*}{\tau}\sigma\left\{p_1(1-k_ {{\bf e}_1})-p_{-1}(1-k_{-{\bf e_{1}}})\right\}k_{{\boldsymbol \lambda}},
\label{systemeg1}
\end{eqnarray}
where $\tilde{L}$ is the operator  
\begin{equation}
\tilde{L}\equiv\sum_\mu A_\mu\nabla_\mu-2d(f+g),
\end{equation}
$\nabla_\mu$ is the discrete gradient defined by $\nabla_\mu f({\boldsymbol \lambda}) \equiv f({\boldsymbol \lambda}+{\bf e}_{ \mu})-f({\boldsymbol \lambda})$, 
and 
\begin{equation}
A_\mu\equiv1+\frac{2d\tau^*}{\tau}p_\mu(1-k_{{\bf e}_{ \mu}}).
\label{defA}
\end{equation}
For  ${\boldsymbol \lambda} = {\bf e_{\nu}}$ 
with $\nu=\{\pm 1,\pm 2,\ldots, \pm d\}$, i.e., the  sites adjacent to the TP, we find
\begin{equation}
2d\tau^*\partial_tk_{{\bf e}_{ \nu}}=(\tilde{L}+A_\nu)k_{{\bf e}_{ \nu}}+2df
\label{systemek2}
\end{equation}
and 
\begin{eqnarray}
2d\tau^*\partial_t\widetilde{g}_{{\bf e}_{ \nu}}&=&(\tilde{L}+A_\nu)\widetilde{g}_{{\bf e}_{ \nu}}+
\frac{2d\tau^*}{\tau}\sigma\left\{ p_1(1-k_{{\bf e}_1})k_{{\bf e}_\nu+{\bf e}_1}\right.\nonumber\\
&-&\left.p_{-1}(1-k_{-{\bf e}_1})k_{{\bf e}_\nu-{\bf e}_1}\right\}-\frac{2d\tau^*}{\tau} p_\nu\widetilde{g}_{{\bf e}_\nu} k_{2{\bf e}_\nu}\nonumber\\
&+&\frac{2d\tau^*}{\tau} p_{-\nu}\widetilde{g}_{-{\bf e}_\nu} k_{{\bf e}_\nu}-\frac{2d\tau^*}{\tau}\sum_{\mu\neq\pm \nu}p_\mu\widetilde{g}_{{\bf e}_{ \mu}}\nabla_\mu k_{{\bf e}_{ \nu}}\nonumber\\
&-&\frac{2d\tau^*}{\tau}\sigma\left\{p_1(1-k_{{\bf e}_1})-p_{-1}(1-k_{-{\bf e}_{1}})\right\}k_{{\bf e}_{ \nu}}.\nonumber\\
\label{systemeg2}
\end{eqnarray}
Note that Eqs.(\ref{systemek2}) and (\ref{systemeg2}) represent 
the boundary conditions for the general evolution equations  (\ref{systemek1}) and (\ref{systemeg1}), imposed on the sites in the
immediate vicinity of the TP. Equations (\ref{systemek1})-(\ref{systemeg2}) together with Eqs.(\ref{Xmean}) and (\ref{dispersion}) constitute
a closed system of equations which suffices for computation of the variance $\langle \delta X_{tr}^2\rangle $ for 
any density and  in arbitrary dimension. These equations also give  access to the cross-correlation functions $\langle \delta X_{tr} \delta\eta_{\boldsymbol \lambda}\rangle$.


\section{Exact Einstein-type relation}

 Our starting point here is the usual Einstein relation $\mu=\beta D$ between the mobility $\mu$ and the diffusion coefficient $D$, which has been shown to hold exactly for the system we study \cite{Komorowski:2005a}. We now use the relations (\ref{Xmean}) and (\ref{dispersion}) in the long time limit to obtain the mobility and the diffusion coefficient in terms of the  density profiles and the correlation functions:
\begin{equation}
\mu=\lim_{E\to0}\frac{\sigma}{E\tau}\{p_1  (1-k_1)-p_{-1} (1-k_{-1})\}
\end{equation}
and 
\begin{eqnarray}
D&=&\lim_{E\to0}\bigg[\frac{\sigma^2}{2d\tau}
\{p_1  (1 -k_1)+p_{-1} (1-k_{-1})\} \nonumber\\
&-&\frac{\sigma}{d\tau}\{p_1  \widetilde{g}_1-p_{-1} \widetilde{g}_{-1}\}\bigg]
\end{eqnarray}
Using Eq(\ref{defp}), as well as the trivial symmetry relations $\widetilde{g}_{-1}(E=0)=-\widetilde{g}_{1}(E=0)$ and $k_{\pm1}=\rho_s\pm\alpha E+o(E)$ where $\alpha$ is a non vanishing constant, we finally get the exact Einstein-type relation of the form
\begin{equation}
\beta \langle \delta X_{tr} \delta\eta_{ {\bf R}_{tr}+ {\bf e}_1} \rangle (E=0) = \lim_{E\to 0}\frac{\langle \eta_{ {\bf R}_{tr}+ {\bf e}_1}\rangle-\rho_s}{E}.
\end{equation}
This equation relates a cross correlation function (between the tracer and the bath)  in the absence of field, to the linear response of the bath itself, and is compatible with the fluctuation dissipation relation described in \cite{Cugliandolo:2011}.

\section{Explicit solution in 1d}

Lengthy but straightforward  calculations  lead to  analytical expressions of the density profiles $k_\lambda$ and cross-correlation functions $\widetilde{g}_{\lambda}$, allowing in particular to determine  the dispersion coefficient 
$K_{tr}=\lim_{t\to\infty}\langle \delta X_{tr}^2\rangle/2t$ (expressions given in appendix \ref{appendixB}). Figure \ref{fig1} shows that this expression of $K_{tr}$  is in excellent agreement with results of numerical simulations for any density, which validates the decoupling approximation  in Eqs. (\ref{prof}) and  (\ref{corr}). We here discuss three important consequences stemming from this expression.

\begin{figure}[h!]
\begin{center}
      \includegraphics[width=8cm]{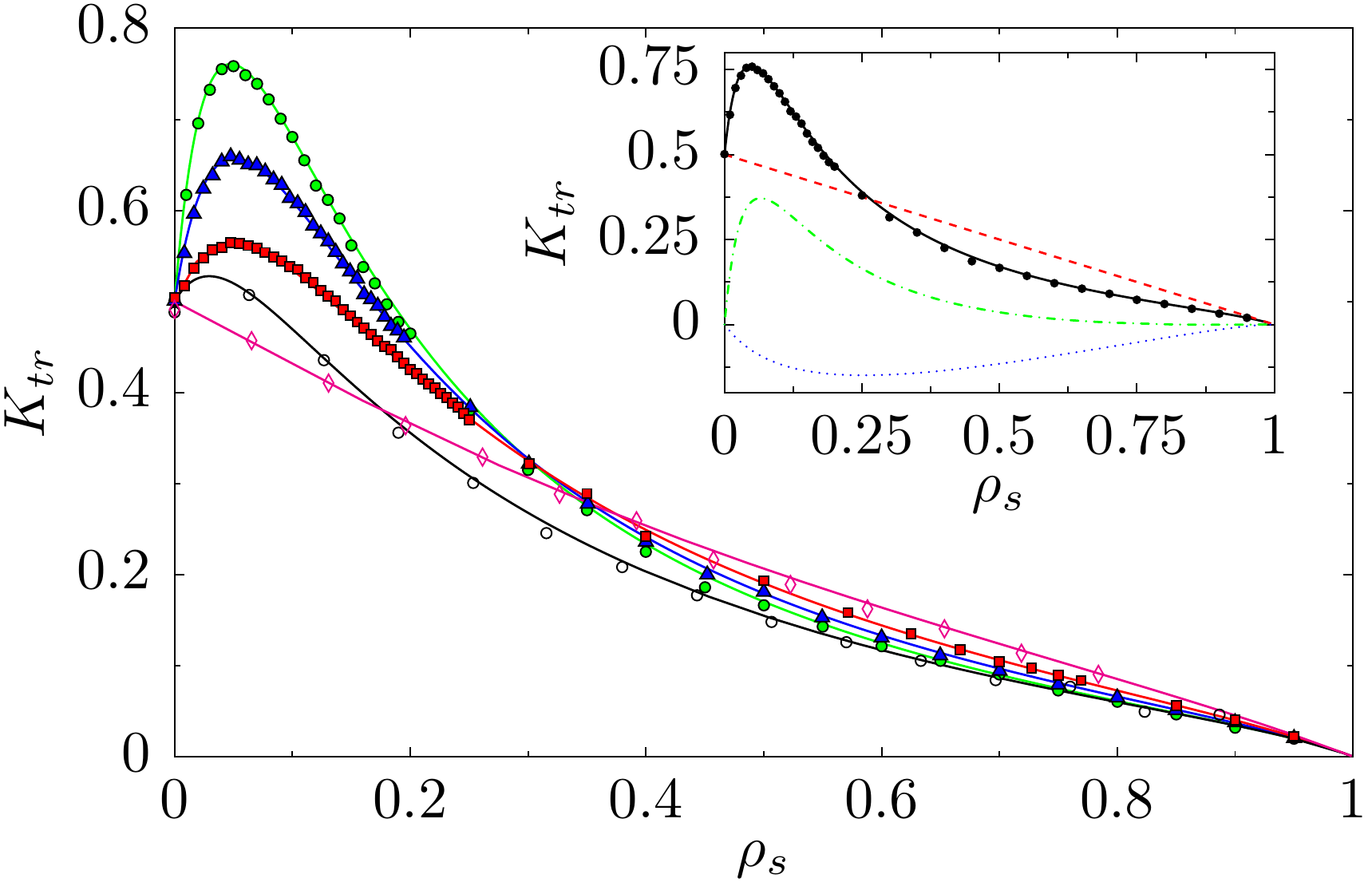}
\end{center}
\caption{Dispersion coefficient as a function of the density  $\rho_s$. Analytical expression (solid lines) vs numerical simulations (symbols) for different biases (filled symbols : $p_1=0.98$ ; empty symbols : $p_1=0.8$) and desorption rates $g$ ($\bullet$, $\circ$ : 0.15 ; $\blacktriangle$ : 0.2 ; $\blacksquare$ : 0.3 ; $\lozenge$ : 0.8). Inset : The different contributions to the dispersion coefficient, as defined in the main text,  as a function of the density $\rho_s$ for  $g=0.15$ : trivial mean-field  (dashed), density (dotted), and cross-correlation (dashdotted) contributions.  The sum of these contributions is the dispersion coefficient (analytical expression represented by a solid line vs numerical simulations denoted by symbols).}
\label{fig1}
\end{figure}

First, in the important limiting situation where no external driving is applied to the system,  we find that 
\begin{eqnarray}
\label{KE0}
K_{tr}(E=0)=\frac{\frac{\sigma^2 (1 - \rho_s)}{2 \tau}}
{1 +\frac{\rho_s\tau^*}{\tau(f+g)} 
\frac{2}{1 + \sqrt{1 +2 (1 + \tau^* (1 - \rho_s)/\tau) /(f+g)}}}.\nonumber\\
\end{eqnarray}
Beyond its own interest, this expression together with the approximate expression for the velocity obtained in \cite{Benichou:1999c} in the linear driving regime using the decoupling approximation (\ref{prof}) shows that the Einstein relation (exact for the system under study \cite{Komorowski:2005a}) is satisfied  under the decoupling approximations, Eqs. (\ref{prof}) and  (\ref{corr}). In other words, the approximations (\ref{prof}) and (\ref{corr}) are compatible with each other, which is a further validation of their applicability.

Second, in the high density limit, $\rho_s\to 1$, the dispersion coefficient obeys
\begin{equation}
K_{tr}\equi{\rho_s\to1}\frac{\sigma^2}{2\tau}\frac{f+\sqrt{f(f+2)}}{f+\sqrt{f(f+2)}+2\tau^*/\tau} (1-\rho_s).
\end{equation}
 At leading order in $1-\rho_s$, $K_{tr}$ is  thus {\it independent} of the amplitude of the driving force. Quite unexpectedly, it turns  out that the  dependence on the driving force of the density and of the correlation contributions to the dispersion coefficient exactly compensate each other in this limit.

Last, our analytical expression shows that, strikingly, $K_{tr}$ is a non-monotonic 
function of the density $\rho_s$,  for values of $g$ smaller than some critical value $g_c$, and has a marked maximum at a certain value of $\rho_s$ (see figure \ref{fig1}). This result is a bit counter-intuitive, 
since one naturally 
expects that the dispersion will be maximal when there are no hard-core bath particles (i.e., when $\rho_s=0$).

\begin{figure}[h!]
\begin{center}
      \includegraphics[width=9cm]{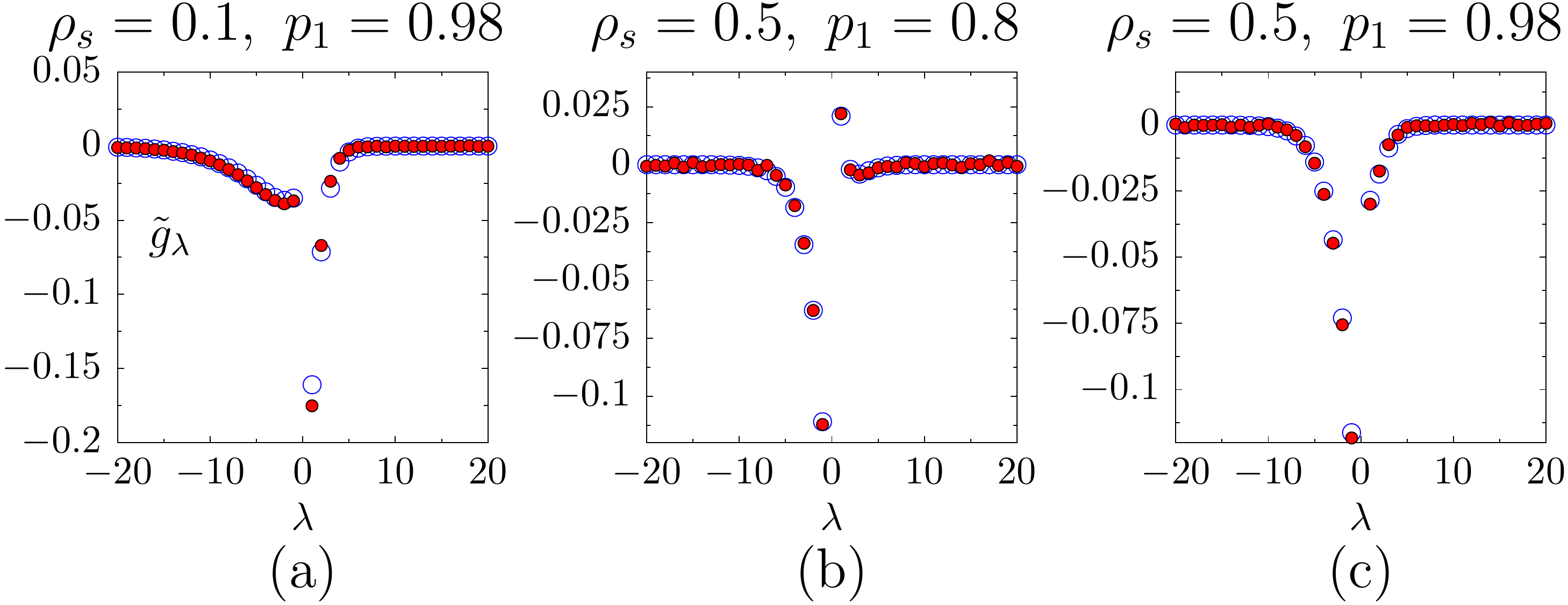}
\end{center}
\caption{Cross-correlation functions $\tilde{g}_\lambda$ as a function of the distance to the tracer $\lambda$ for different values of the density $\rho_s$ and of the bias $p_1$. Analytical expressions (empty symbols) vs numerical simulations (filled symbols). (a) $f=0.033$, $g=0.3$ ; (b) and (c) $f=0.3$, $g=0.3$.}\label{ggprofile}
\end{figure}

To gain some insight into this non trivial behavior, we represent  (see inset of fig \ref{fig1}) three different contributions to the dispersion coefficient defined above (trivial mean-field, density, and cross-correlation contributions). As can be immediately seen, the non monotonic behavior of the variance  originates from the non-monotonic behavior  with $\rho_s$ of the cross-correlation function $\widetilde{g}_1$ in front of the tracer. These cross-correlations functions $\widetilde{g}_{\lambda}$, which    provide  a quantitative measure of the correlations between the fluctuations of the tracer position and the occupation numbers of the bath particles,  are thus crucial in the problem. Actually, they  display other non trivial and important behaviours (see fig. \ref{ggprofile}) revealed by our analytical expressions.

We first stress that the sign itself of these quantities is not trivial. While the analytical expressions allow us to show that, past the tracer,  $\widetilde{g}_{\lambda}$ is always negative and corresponds to an anti-correlation of the fluctuations, it is found that $\widetilde{g}_{1}$ is positive in the high density limit 
\begin{equation}
\widetilde{g}_{1}\equi{\rho_s\to1}\frac{2p_{-1}\sigma(1-\rho_s)}{2+[f+\sqrt{f(2+f)}]\tau/\tau^*}>0
\end{equation}
or in the absence of a driving force
\begin{equation}
\widetilde{g}_{1}(E=0)\sim\frac{\sigma(1-\rho_s)}{2}-\frac{\tau}{\sigma}K_{tr}(E=0)>0
\end{equation}
where $K_{tr}(E=0)$ is given by Eq.(\ref{KE0}), but can change sign otherwise. In other words, depending on the amplitude of the driving force and the density of the bath particles, positive or negative correlations between the fluctuations of the tracer position and the occupation numbers of the bath particles can take place.
In addition, the monotonicity with the distance of these cross-correlation functions is also non-trivial -- past (see fig. \ref{ggprofile}(a)) and in front of (see fig. \ref{ggprofile}(b)) the tracer -- which further mirrors the complexity of these  correlations.

\section{Conclusion}

In conclusion,  we have studied the dynamics of a driven tracer  in a bath of hard-core  particles in the presence of adsorption/desorption processes. For this   minimal model of active microrheology, we have developed an analytical framework to go beyond the standard force-velocity relation 
   and quantitatively analyzed the fluctuations of the position of the tracer and their correlations with the occupation number of the bath particles at any position, for {\it any} density of the bath particles. This  approach allowed us to obtain an exact Einstein-type relation between these correlations in the absence of a driving force and the density profiles of the bath particles in the linear  driving regime. In the one-dimensional case we also provided an approximate but accurate explicit expression of the variance of the position of the tracer   and  shown in particular that it can behave non-monotically  with the density of bath particles: counter-intuitively, increasing the density of hard-core particles can significantly increase the dispersion of the tracer position. We have shown that subtle  correlations between the tracer position and the occupation of the bath particles past and in front of the tracer are responsible for this non trivial behavior.  
   Altogether, our results put forward  cross correlations as    observables  to analyze active microrheology data and predict a striking  non monotonic behavior of the variance of the position of the tracer with respect to the density of the bath, that could in principle be observed experimentally.
In a broader context,  Equations (\ref{systemek1}) to (\ref{systemeg2}) together with Eqs.(\ref{Xmean}) and (\ref{dispersion}) provide a general framework to investigate important extensions   that include the determination of  explicit solutions in higher dimensions, in confined geometries and in non-stationary regimes.

\section*{Acknowledgment}

O.B. acknowledges support from the European Research Council (Grant No. FPTOpt-277998).

\appendix

\section{Master equation}
\label{appendixA}

Let $\eta\equiv
\{\eta_{\bf R}\}$ denote the entire set of the occupation variables, which defines the instantaneous
configuration of the adsorbed particles  at the 
lattice at time moment $t$. Next, let 
$P({\bf R}_{tr},\eta;t)$ stand for the joint probability of finding  
at time $t$ the TP at the site ${\bf R}_{tr}$
and all adsorbed particles in the configuration $\eta$.
Then, denoting as $\eta^{{\bf r},\nu}$ a
configuration obtained from $\eta$ by the Kawasaki-type exchange of the occupation
variables of two neighboring 
sites ${\bf r}$ and ${\bf r+e}_{ \nu}$,
and as 
${\hat \eta}^{\bf r}$  a configuration obtained from 
the original  $\eta$ by
the replacement $\eta_{\bf r} \to 1-\eta_{\bf r}$, which corresponds to the
Glauber-type flip of the occupation variable due to 
the adsorption/desorption 
events, we have that 
the time evolution of the configuration
probability $P({\bf R}_{tr},\eta;t)$ obeys the following master equation~:

\begin{eqnarray}
&&2d\tau^*\partial_tP({\bf R}_{tr},\eta;t)=\nonumber\\
&&\sum_{\mu=1}^d\;\sum_{{\bf r}\neq{\bf R}_{tr}-{\bf e}_{ \mu},{\bf R}_{tr}}  \; 
\Big\{ P({\bf R}_{tr},\eta^{{\bf r},\mu};t)-P({\bf R}_{tr},\eta;t)\Big\}\nonumber\\
&&+\frac{2d\tau^*}{\tau}\sum_{\mu}p_\mu\Big\{\left(1-\eta_{{\bf R}_{tr}}\right)P({\bf R}_{tr}-{\bf e}_{ \mu},\eta;t)\nonumber\\
&&-\left(1-\eta_{{\bf R}_{tr}+{\bf e}_{ \mu}}\right)P({\bf R}_{tr},\eta;t)\Big\}\nonumber\\
&&+2dg\sum_{{\bf r}\neq {\bf R}_{tr}} \;\Big\{\left(1-\eta_{\bf r}\right)P({\bf R}_{tr},
\hat{\eta}^{{\bf r}};t)-\eta_{\bf r}P({\bf R}_{tr},\eta;t)\Big\}\nonumber\\
&&+2df\sum_{{\bf r}\neq{\bf R}_{tr}} \;\Big\{\eta_{\bf r}P({\bf R}_{tr},\hat{\eta}^{{\bf r}};t)
-\left(1-\eta_{\bf r}\right)P({\bf R}_{tr},\eta;t)\Big\}.\nonumber\\
\label{eqmaitresse}
\end{eqnarray}

Note that the time evolution of the first moment $\langle X_{tr}(t)\rangle$ (Eq. \ref{Xmean}) can be obtained by multiplying both sides of Eq. (\ref{eqmaitresse}) by $({\bf R}_{tr} \cdot {\bf e}_1)$ and summing over all possible configurations $({\bf R}_{tr},\eta)$.

Eqs. \ref{systemek1} and \ref{systemek2} (respectively \ref{systemeg1} and \ref{systemeg2}) are obtained by multiplying both sides of Eq. (\ref{eqmaitresse}) by $\eta({\bf R}_{tr}+{\boldsymbol \lambda})$ (respectively $X_{tr} \eta({\bf R}_{tr}+{\boldsymbol \lambda}) $), summing over all possible configurations, and using the  approximation \ref{prof} (respectively \ref{corr}).

\section{Explicit expression of $K_{tr}$}
\label{appendixB}

\subsection{Determination of $k_1$ and $k_{-1}$}

According to Eq. \ref{dispersion}, and defining $K_{tr}=\lim_{t\to\infty}\langle \delta X_{tr}^2\rangle/2t$, 
\begin{eqnarray}
\label{Ktr}
&&K_{tr}  = \frac{\sigma^2}{2\tau}(1-\rho_s) - \frac{\sigma^2}{2\tau} \left[p_1(k_1-\rho_s)\right.\nonumber\\
&&\left.+p_{-1}(k_{-1}-\rho_s)\right]-\frac{\sigma}{\tau}\left(p_1\widetilde{g}_{1}-p_{-1}\widetilde{g}_{-1}\right).\nonumber\\
\end{eqnarray}
Thus, the computation of $k_{\pm1}$ and $\tilde{g}_{\pm1}$ from Eqs. \ref{systemek1}, \ref{systemek2}, \ref{systemeg1}, and \ref{systemeg2} gives the expression of $K_{tr}$.\\

For one-dimensional lattices, a general solution of
 Eqs. \ref{systemek1} and \ref{systemek2} has the following form:
 
\begin{equation}
\label{kcases}
k_{n} \equiv k(\lambda)=
\begin{cases}
   \rho_s + K_{+} r_1^n    & \text{for $n>0$}, \\
  \rho_s + K_{-} r_2^n    & \text{for $n<0$}.
\end{cases}
\end{equation}

where
\begin{eqnarray}
\label{r1}
r_{\substack{2\\1}} &=& \frac{1}{2A_1}\bigg( A_{1} + A_{-1} + 2 (f + g) \nonumber \\
&\pm& \left.\sqrt{\Big(A_{1} + A_{-1} + 2 (f + g)\Big)^2 - 4 A_{1} A_{-1}}\right), \nonumber \\
\end{eqnarray} 

while the amplitudes $K_{\pm}$ are given respectively by
\begin{eqnarray}
\label{Kp}
K_{+} &=& \rho_s \frac{A_{1} - A_{-1}}{A_{-1} - A_{1} r_1}, \\
\label{Km} 
K_{-} &=& \rho_s \frac{A_{1} - A_{-1}}{A_{-1}/r_2 - A_{1} } .
\end{eqnarray}

Now, we are in position to obtain a system of two closed-form non-linear equations determining implicitly the unknown 
parameters $A_{1}$ and $A_{-1}$, which will allow us to compute
$k_1$ and $k_{-1}$.
Substituting Eqs. \ref{kcases}, \ref{Kp} and \ref{Km} into Eq. \ref{defA}, we find 
\begin{eqnarray}
\label{Ap}
A_{1} &=& 1 + \frac{2p_1 \tau^*}{\tau} \Big[1 - \rho_s - \rho_s \frac{A_{1} - A_{-1}}{A_{-1} /r_1 - A_{1}},
\Big]\\
\label{Am}
A_{-1} &=& 1 + \frac{2p_{-1} \tau^*}{\tau} \Big[1 - \rho_s - \rho_s \frac{A_{1} - A_{-1}}{A_{-1}  - A_{1} r_2}
\Big].
\end{eqnarray}

\subsection{Determination of $\tilde{g}_1$ and $\tilde{g}_{-1}$}

Using  Eqs. \ref{systemeg1} and \ref{systemeg2} and the expressions \ref{kcases} of $k_n$, the general equations satisfied by $\widetilde{g}$ become
\begin{eqnarray}
&&A_1(\widetilde{g}_{n+1}-\widetilde{g}_{n})+A_{-1}(\widetilde{g}_{n-1}-\widetilde{g}_{n})-2(f+g)\widetilde{g}_{n}\nonumber\\
&+&\frac{2\tau^*}{\tau}\sigma\left\{p_1K_+r_1^n\left(1-\rho_s-K_+r_1-\frac{\widetilde{g}_{1}}{\sigma}\right)(r_1-1)\right.\nonumber\\
&-&\left.p_{-1}K_+r_1^n\left(1-\rho_s-K_+r_1+\frac{\widetilde{g}_{-1}}{\sigma}\right)(r_1^{-1}-1)\right\} =0\nonumber\\
\end{eqnarray}
 for $n>1$, 
  \begin{eqnarray}
  &&A_1(\widetilde{g}_{n+1}-\widetilde{g}_{n})+A_{-1}(\widetilde{g}_{n-1}-\widetilde{g}_{n})-2(f+g)\widetilde{g}_{n}\nonumber\\
&+&\frac{2\tau^*}{\tau}\sigma\left\{p_1K_-r_2^n\left(1-\rho_s-K_+r_1-\frac{\widetilde{g}_{1}}{\sigma}\right)(r_2-1)\right.\nonumber\\
&-&\left.p_{-1}K_-r_2^n\left(1-\rho_s-K_+r_1+\frac{\widetilde{g}_{-1}}{\sigma}\right)(r_2^{-1}-1)\right\} =0\nonumber\\
\end{eqnarray}
for $n<-1$,
\begin{eqnarray}
\label{CL1}
&&A_1\widetilde{g}_{2}-\widetilde{g}_{1}\left(A_{-1}+2(f+g)+\frac{2\tau^*}{\tau}p_1(\rho_s+K_+r_1^2)\right)\nonumber\\
&+&\frac{2\tau^*}{\tau}p_{-1}(\rho_s+K_+r_1)\widetilde{g}_{-1}\nonumber\\
&=&-\frac{2\tau^*}{\tau}\sigma p_1(1-\rho_s-K_+r_1)(\rho_s+K_+r_1^2)\nonumber\\
&+&\frac{2\tau^*}{\tau}\sigma [p_1(1-\rho_s-K_+r_1)\nonumber\\
&-&p_{-1}(1-\rho_s-K_-r_2^{-1})](\rho_s+K_+r_1),
\end{eqnarray}
and 
\begin{eqnarray}
\label{CL2}
&&A_{-1}\widetilde{g}_{-2}-\widetilde{g}_{-1}\left(A_{1}+2(f+g)+\frac{2\tau^*}{\tau}p_{-1}(\rho_s+K_-r_2^{-2})\right)\nonumber\\
&+&\frac{2\tau^*}{\tau}p_{1}(\rho_s+K_-r_2^{-1})\widetilde{g}_{1}\nonumber\\
&=&\frac{2\tau^*}{\tau}\sigma p_{-1}(1-\rho_s-K_-r_2^{-1})(\rho_s+K_-r_2^{-2})\nonumber\\
&+&\frac{2\tau^*}{\tau}\sigma [p_1(1-\rho_s-K_+r_1)\nonumber\\
&-&p_{-1}(1-\rho_s-K_-r_2^{-1})](\rho_s+K_-r_2^{-1}).\nonumber\\
\end{eqnarray}

The general solution is written
\begin{equation}
\label{gcases}
\widetilde{g}_n=
\begin{cases}
  \alpha r_1^n-\dfrac{V}{A_1r_1-A_{-1}r_1^{-1}}nr_1^n    & \text{for $n>0$}, \\
   \beta r_2^n-\dfrac{W}{A_1r_2-A_{-1}r_2^{-1}}nr_2^n   & \text{for $n<0$}.
\end{cases}
\end{equation}
where $\alpha$ and $\beta$ are constants to be determined, 
\begin{eqnarray}
V&\equiv&K_+\frac{2\tau^*}{\tau}\sigma\left\{p_1\left(1-\rho_s-K_+r_1-\frac{\widetilde{g}_{1}}{\sigma}\right)(r_1-1)\right.\nonumber\\
&-&\left.p_{-1}\left(1-\rho_s-K_+r_1+\frac{\widetilde{g}_{-1}}{\sigma}\right)(r_1^{-1}-1)\right\}
\end{eqnarray}
and
\begin{eqnarray}
W&\equiv&K_-\frac{2\tau^*}{\tau}\sigma\left\{p_1\left(1-\rho_s-K_+r_1-\frac{\widetilde{g}_{1}}{\sigma}\right)(r_2-1)\right.\nonumber\\
&-&\left.p_{-1}\left(1-\rho_s-K_+r_1+\frac{\widetilde{g}_{-1}}{\sigma}\right)(r_2^{-1}-1)\right\}.
\end{eqnarray}

Substituting Eq. \ref{gcases} into Eqs. \ref{CL1} and \ref{CL2} on the one hand, and writing 
Eq. \ref{gcases} for $n=1$ and $n=-1$ on the other hand, leads to a linear system of four equations satisfied by the four unknowns
$\alpha$, $\beta$, $\widetilde{g}_{1}$ and $\widetilde{g}_{-1}$. It is finally found that:
\begin{equation}
\label{g1}
\widetilde{g}_{1}=\frac{CE-BF}{AE-BD} \;\;{\rm and }\;\;\widetilde{g}_{-1}=\frac{AF-CD}{AE-BD},
\end{equation}
where
\begin{eqnarray}
A&\equiv&\frac{2\tau^*}{\tau}\sigma p_1K_+(r_1-1)\frac{A_1r_1^2}{A_1r_1-A_{-1}r_1^{-1}}+A_1r_1\nonumber\\
&-&\left(A_{-1}+2(f+g)+\frac{2\tau^*}{\tau}p_1(\rho_s+K_+r_1^2)\right),
\end{eqnarray}
\begin{eqnarray}
B&\equiv&\frac{2\tau^*}{\tau}\sigma p_{-1}K_+(r_1^{-1}-1)\frac{A_1r_1^2}{A_1r_1-A_{-1}r_1^{-1}}\nonumber\\
&+&\frac{2\tau^*}{\tau}p_{-1}(\rho_s+K_+r_1),
\end{eqnarray}
\begin{eqnarray}
C&\equiv&\frac{2\tau^*}{\tau}\sigma (p_1(1-\rho_s-K_+r_1)\nonumber\\
&-&p_{-1}(1-\rho_s-K_-r_2^{-1}))(\rho_s+K_+r_1)\nonumber\\
&-&\frac{2\tau^*}{\tau}\sigma p_1(1-\rho_s-K_+r_1)(\rho_s+K_+r_1^2)\nonumber\\
&+&\frac{2\tau^*}{\tau}\sigma\bigg[p_1(1-\rho_s-K_+r_1)K_+r_1\nonumber\\
&-&\left.p_{-1}\left(1-\rho_s-\frac{K_-}{r_2}\right)\frac{K_+}{r_1}\right]\frac{A_1r_1^2}{A_1r_1-A_{-1}r_1^{-1}},\nonumber\\
\end{eqnarray}
\begin{eqnarray}
D&\equiv&-\frac{2\tau^*}{\tau}\sigma p_{1}K_-(r_2-1)\frac{A_{-1}r_2^{-2}}{A_1r_2-A_{-1}r_2^{-1}}\nonumber\\
&+&\frac{2\tau^*}{\tau}p_{1}(\rho_s+K_-r_2^{-1}),
\end{eqnarray}
\begin{eqnarray}
E&\equiv&-\frac{2\tau^*}{\tau}\sigma p_{-1}K_-(r_2^{-1}-1)\frac{A_{-1}r_2^{-2}}{A_1r_2-A_{-1}r_2^{-1}}+A_{-1}r_2^{-1}\nonumber\\
&-&\left(A_{1}+2(f+g)+\frac{2\tau^*}{\tau}p_{-1}(\rho_s+K_-r_2^{-2})\right)
\end{eqnarray}
and
\begin{eqnarray}
F&\equiv&\frac{2\tau^*}{\tau}\sigma (p_1(1-\rho_s-K_+r_1)\nonumber\\
&-&p_{-1}(1-\rho_s-K_-r_2^{-1}))(\rho_s+K_-r_2^{-1})\nonumber\\
&+&\frac{2\tau^*}{\tau}\sigma p_{-1}(1-\rho_s-K_-r_2^{-1})(\rho_s+K_-r_2^{-2})\nonumber\\
&-&\frac{2\tau^*}{\tau}\sigma\bigg[(p_1(1-\rho_s-K_+r_1)K_-r_2\nonumber\\
&-&\left.p_{-1}\left(1-\rho_s-\frac{K_-}{r_2}\right)\frac{K_-}{r_2}\right]\frac{A_{-1}r_2^{-2}}{A_1r_2-A_{-1}r_2^{-1}},\nonumber\\
\end{eqnarray}
where $K_+$, $K_-$, $A_1$, and $A_{-1}$ have been determined in the previous section.


\begin{thebibliography}{22}
\expandafter\ifx\csname natexlab\endcsname\relax\def\natexlab#1{#1}\fi
\expandafter\ifx\csname bibnamefont\endcsname\relax
  \def\bibnamefont#1{#1}\fi
\expandafter\ifx\csname bibfnamefont\endcsname\relax
  \def\bibfnamefont#1{#1}\fi
\expandafter\ifx\csname citenamefont\endcsname\relax
  \def\citenamefont#1{#1}\fi
\expandafter\ifx\csname url\endcsname\relax
  \def\url#1{\texttt{#1}}\fi
\expandafter\ifx\csname urlprefix\endcsname\relax\def\urlprefix{URL }\fi
\providecommand{\bibinfo}[2]{#2}
\providecommand{\eprint}[2][]{\url{#2}}

\bibitem[{\citenamefont{Wilson and Poon}(2011)}]{Wilson:2011a}
\bibinfo{author}{\bibfnamefont{L.~G.} \bibnamefont{Wilson}} \bibnamefont{and}
  \bibinfo{author}{\bibfnamefont{W.~C.~K.} \bibnamefont{Poon}},
  \bibinfo{journal}{PhysChemChemPhys}
  \textbf{\bibinfo{volume}{13}}, \bibinfo{pages}{10617} (\bibinfo{year}{2011}).

\bibitem[{\citenamefont{Wilson et~al.}(2011)\citenamefont{Wilson, Harrison,
  Poon, and Puertas}}]{Wilson:2011}
\bibinfo{author}{\bibfnamefont{L.~G.} \bibnamefont{Wilson}},
  \bibinfo{author}{\bibfnamefont{A.~W.} \bibnamefont{Harrison}},
  \bibinfo{author}{\bibfnamefont{W.~C.~K.} \bibnamefont{Poon}},
  \bibnamefont{and} \bibinfo{author}{\bibfnamefont{A.~M.}
  \bibnamefont{Puertas}}, \bibinfo{journal}{Europhys. Lett.}
  \textbf{\bibinfo{volume}{93}}, \bibinfo{pages}{58007} (\bibinfo{year}{2011}).

\bibitem[{\citenamefont{Winter et~al.}(2012)\citenamefont{Winter, Horbach,
  Virnau, and Binder}}]{Winter:2012}
\bibinfo{author}{\bibfnamefont{D.}~\bibnamefont{Winter}},
  \bibinfo{author}{\bibfnamefont{J.}~\bibnamefont{Horbach}},
  \bibinfo{author}{\bibfnamefont{P.}~\bibnamefont{Virnau}}, \bibnamefont{and}
  \bibinfo{author}{\bibfnamefont{K.}~\bibnamefont{Binder}},
  \bibinfo{journal}{Phys. Rev. Lett.} \textbf{\bibinfo{volume}{108}},
  \bibinfo{pages}{028303} (\bibinfo{year}{2012}).

\bibitem[{\citenamefont{{Schroer} and {Heuer}}(2012)}]{2012arXiv1210.1081S}
\bibinfo{author}{\bibfnamefont{C.~F.~E.} \bibnamefont{{Schroer}}}
  \bibnamefont{and} \bibinfo{author}{\bibfnamefont{A.}~\bibnamefont{{Heuer}}},
 \bibinfo{journal}{Phys. Rev. Lett.} \textbf{\bibinfo{volume}{110}},
  \bibinfo{pages}{067801} (\bibinfo{year}{2013}).

\bibitem[{\citenamefont{{B{\'e}nichou}
  et~al.}(2012)\citenamefont{{B{\'e}nichou}, {Mej{\'{\i}}a-Monasterio}, and
  {Oshanin}}}]{2012arXiv1211.0674B}
\bibinfo{author}{\bibfnamefont{O.}~\bibnamefont{{B{\'e}nichou}}},
  \bibinfo{author}{\bibfnamefont{C.}~\bibnamefont{{Mej{\'{\i}}a-Monasterio}}},
  \bibnamefont{and}
  \bibinfo{author}{\bibfnamefont{G.}~\bibnamefont{{Oshanin}}},
  \bibinfo{journal}{Phys. Rev. E} \textbf{\bibinfo{volume}{87}},
  \bibinfo{pages}{020103} (\bibinfo{year}{2013}).

\bibitem[{\citenamefont{Harrer et~al.}(2012)\citenamefont{Harrer, Winter,
  Horbach, Fuchs, and Voigtmann}}]{Harrer:2012}
\bibinfo{author}{\bibfnamefont{C.~J.} \bibnamefont{Harrer}},
  \bibinfo{author}{\bibfnamefont{D.}~\bibnamefont{Winter}},
  \bibinfo{author}{\bibfnamefont{J.}~\bibnamefont{Horbach}},
  \bibinfo{author}{\bibfnamefont{M.}~\bibnamefont{Fuchs}}, \bibnamefont{and}
  \bibinfo{author}{\bibfnamefont{T.}~\bibnamefont{Voigtmann}},
  \bibinfo{journal}{J. Phys.: Condens. Matter}
  \textbf{\bibinfo{volume}{24}}, \bibinfo{pages}{464105}
  (\bibinfo{year}{2012}).

\bibitem[{\citenamefont{Marini Bettolo~Marconi
  et~al.}(2008)\citenamefont{Marini Bettolo~Marconi, Puglisi, Rondoni, and
  Vulpiani}}]{Marconi:2008}
\bibinfo{author}{\bibfnamefont{U.}~\bibnamefont{Marini Bettolo~Marconi}},
  \bibinfo{author}{\bibfnamefont{A.}~\bibnamefont{Puglisi}},
  \bibinfo{author}{\bibfnamefont{L.}~\bibnamefont{Rondoni}}, \bibnamefont{and}
  \bibinfo{author}{\bibfnamefont{A.}~\bibnamefont{Vulpiani}},
  \bibinfo{journal}{Phys. Rep.} \textbf{\bibinfo{volume}{461}},
  \bibinfo{pages}{111} (\bibinfo{year}{2008}).

\bibitem[{\citenamefont{Cugliandolo}(2011)}]{Cugliandolo:2011}
\bibinfo{author}{\bibfnamefont{L.~F.} \bibnamefont{Cugliandolo}},
  \bibinfo{journal}{J. Phys. A}
  \textbf{\bibinfo{volume}{44}}, \bibinfo{pages}{483001}
  (\bibinfo{year}{2011}).

\bibitem[{\citenamefont{Gradenigo et~al.}(2012)\citenamefont{Gradenigo,
  Puglisi, Sarracino, Villamaina, and Vulpiani}}]{vulp}
\bibinfo{author}{\bibfnamefont{G.}~\bibnamefont{Gradenigo}},
  \bibinfo{author}{\bibfnamefont{A.}~\bibnamefont{Puglisi}},
  \bibinfo{author}{\bibfnamefont{A.}~\bibnamefont{Sarracino}},
  \bibinfo{author}{\bibfnamefont{D.}~\bibnamefont{Villamaina}},
  \bibnamefont{and} \bibinfo{author}{\bibfnamefont{A.}~\bibnamefont{Vulpiani}},
  \emph{\bibinfo{title}{Out-of-equilibrium generalized fluctuation-dissipation
  relations}} (\bibinfo{publisher}{Chapter in the book: R.Klages, W.Just,
  C.Jarzynski (Eds.), Nonequilibrium Statistical Physics of Small Systems:
  Fluctuation Relations and Beyond (Wiley-VCH, Weinheim, 2012; ISBN
  978-3-527-41094-1)}, \bibinfo{year}{2012}).

\bibitem[{\citenamefont{Lizana et~al.}(2010)\citenamefont{Lizana,
  Ambj{\"o}rnsson, Taloni, Barkai, and Lomholt}}]{Lizana:2010}
\bibinfo{author}{\bibfnamefont{L.}~\bibnamefont{Lizana}},
  \bibinfo{author}{\bibfnamefont{T.}~\bibnamefont{Ambj{\"o}rnsson}},
  \bibinfo{author}{\bibfnamefont{A.}~\bibnamefont{Taloni}},
  \bibinfo{author}{\bibfnamefont{E.}~\bibnamefont{Barkai}}, \bibnamefont{and}
  \bibinfo{author}{\bibfnamefont{M.~A.} \bibnamefont{Lomholt}},
  \bibinfo{journal}{Phys. Rev. E} \textbf{\bibinfo{volume}{81}},
  \bibinfo{pages}{051118} (\bibinfo{year}{2010}).

\bibitem[{\citenamefont{Harris}(1965)}]{Harris:1965}
\bibinfo{author}{\bibfnamefont{T.~E.} \bibnamefont{Harris}},
  \bibinfo{journal}{J. of Appl. Prob.}
  \textbf{\bibinfo{volume}{2}}, \bibinfo{pages}{323} (\bibinfo{year}{1965}).

\bibitem[{\citenamefont{Druger et~al.}(1983)\citenamefont{Druger, Nitzan, and
  Ratner}}]{Druger:1983a}
\bibinfo{author}{\bibfnamefont{S.~D.} \bibnamefont{Druger}},
  \bibinfo{author}{\bibfnamefont{A.}~\bibnamefont{Nitzan}}, \bibnamefont{and}
  \bibinfo{author}{\bibfnamefont{M.~A.} \bibnamefont{Ratner}},
  \bibinfo{journal}{J. Chem. Phys.}
  \textbf{\bibinfo{volume}{79}}, \bibinfo{pages}{3133} (\bibinfo{year}{1983}).

\bibitem[{\citenamefont{B{\'e}nichou
  et~al.}(2000{\natexlab{a}})\citenamefont{B{\'e}nichou, Klafter, Moreau, and
  Oshanin}}]{Benichou:2000la}
\bibinfo{author}{\bibfnamefont{O.}~\bibnamefont{B{\'e}nichou}},
  \bibinfo{author}{\bibfnamefont{J.}~\bibnamefont{Klafter}},
  \bibinfo{author}{\bibfnamefont{M.}~\bibnamefont{Moreau}}, \bibnamefont{and}
  \bibinfo{author}{\bibfnamefont{G.}~\bibnamefont{Oshanin}},
  \bibinfo{journal}{Phys. Rev. E} \textbf{\bibinfo{volume}{62}}, \bibinfo{pages}{3327}
  (\bibinfo{year}{2000}{\natexlab{a}}).

\bibitem[{\citenamefont{Chou et~al.}(2011)\citenamefont{Chou, Mallick, and
  Zia}}]{Chou:2011}
\bibinfo{author}{\bibfnamefont{T.}~\bibnamefont{Chou}},
  \bibinfo{author}{\bibfnamefont{K.}~\bibnamefont{Mallick}}, \bibnamefont{and}
  \bibinfo{author}{\bibfnamefont{R.~K.~P.} \bibnamefont{Zia}},
  \bibinfo{journal}{Rep. on Prog. in Phys.}
  \textbf{\bibinfo{volume}{74}}, \bibinfo{pages}{116601}
  (\bibinfo{year}{2011}).

\bibitem[{\citenamefont{Parmeggiani et~al.}(2003)\citenamefont{Parmeggiani,
  Franosch, and Frey}}]{Parmeggiani:2003}
\bibinfo{author}{\bibfnamefont{A.}~\bibnamefont{Parmeggiani}},
  \bibinfo{author}{\bibfnamefont{T.}~\bibnamefont{Franosch}}, \bibnamefont{and}
  \bibinfo{author}{\bibfnamefont{E.}~\bibnamefont{Frey}},
  \bibinfo{journal}{Phys. Rev. Lett.} \textbf{\bibinfo{volume}{90}},
  \bibinfo{pages}{086601} (\bibinfo{year}{2003}).

\bibitem[{\citenamefont{Burlatsky et~al.}(1992)\citenamefont{Burlatsky,
  Oshanin, Mogutov, and Moreau}}]{Burlatsky:1992a}
\bibinfo{author}{\bibfnamefont{S.~F.} \bibnamefont{Burlatsky}},
  \bibinfo{author}{\bibfnamefont{G.}~\bibnamefont{Oshanin}},
  \bibinfo{author}{\bibfnamefont{A.}~\bibnamefont{Mogutov}}, \bibnamefont{and}
  \bibinfo{author}{\bibfnamefont{M.}~\bibnamefont{Moreau}},
  \bibinfo{journal}{Phys. Lett. A} \textbf{\bibinfo{volume}{166}},
  \bibinfo{pages}{230} (\bibinfo{year}{1992}).

\bibitem[{\citenamefont{Burlatsky et~al.}(1996)\citenamefont{Burlatsky,
  Oshanin, Moreau, and Reinhardt}}]{Burlatsky:1996b}
\bibinfo{author}{\bibfnamefont{S.~F.} \bibnamefont{Burlatsky}},
  \bibinfo{author}{\bibfnamefont{G.}~\bibnamefont{Oshanin}},
  \bibinfo{author}{\bibfnamefont{M.}~\bibnamefont{Moreau}}, \bibnamefont{and}
  \bibinfo{author}{\bibfnamefont{W.~P.} \bibnamefont{Reinhardt}},
  \bibinfo{journal}{Phys. Rev. E} \textbf{\bibinfo{volume}{54}},
  \bibinfo{pages}{3165} (\bibinfo{year}{1996}).

\bibitem[{\citenamefont{B{\'e}nichou et~al.}(1999)\citenamefont{B{\'e}nichou,
  Cazabat, Lemarchand, Moreau, and Oshanin}}]{Benichou:1999c}
\bibinfo{author}{\bibfnamefont{O.}~\bibnamefont{B{\'e}nichou}},
  \bibinfo{author}{\bibfnamefont{A.~M.} \bibnamefont{Cazabat}},
  \bibinfo{author}{\bibfnamefont{A.}~\bibnamefont{Lemarchand}},
  \bibinfo{author}{\bibfnamefont{M.}~\bibnamefont{Moreau}}, \bibnamefont{and}
  \bibinfo{author}{\bibfnamefont{G.}~\bibnamefont{Oshanin}},
  \bibinfo{journal}{J. of Stat. Phys.}
  \textbf{\bibinfo{volume}{97}}, \bibinfo{pages}{351} (\bibinfo{year}{1999}).

\bibitem[{\citenamefont{B{\'e}nichou
  et~al.}(2000{\natexlab{b}})\citenamefont{B{\'e}nichou, Cazabat, De~Coninck,
  Moreau, and Oshanin}}]{Benichou:2000gd}
\bibinfo{author}{\bibfnamefont{O.}~\bibnamefont{B{\'e}nichou}},
  \bibinfo{author}{\bibfnamefont{A.}~\bibnamefont{Cazabat}},
  \bibinfo{author}{\bibfnamefont{J.}~\bibnamefont{De~Coninck}},
  \bibinfo{author}{\bibfnamefont{M.}~\bibnamefont{Moreau}}, \bibnamefont{and}
  \bibinfo{author}{\bibfnamefont{G.}~\bibnamefont{Oshanin}},
  \bibinfo{journal}{Phys. Rev. Lett.} \textbf{\bibinfo{volume}{84}},
  \bibinfo{pages}{511} (\bibinfo{year}{2000}{\natexlab{b}}).

\bibitem[{\citenamefont{B{\'e}nichou et~al.}(2001)\citenamefont{B{\'e}nichou,
  Cazabat, De~Coninck, Moreau, and Oshanin}}]{Benichou:2001a}
\bibinfo{author}{\bibfnamefont{O.}~\bibnamefont{B{\'e}nichou}},
  \bibinfo{author}{\bibfnamefont{A.~M.} \bibnamefont{Cazabat}},
  \bibinfo{author}{\bibfnamefont{J.}~\bibnamefont{De~Coninck}},
  \bibinfo{author}{\bibfnamefont{M.}~\bibnamefont{Moreau}}, \bibnamefont{and}
  \bibinfo{author}{\bibfnamefont{G.}~\bibnamefont{Oshanin}},
  \bibinfo{journal}{Phys. Rev. B} \textbf{\bibinfo{volume}{63}},
  \bibinfo{pages}{235413} (\bibinfo{year}{2001}).

\bibitem[{\citenamefont{Landim et~al.}(1998)\citenamefont{Landim, Olla, and
  Volchan}}]{Landim:1998a}
\bibinfo{author}{\bibfnamefont{C.}~\bibnamefont{Landim}},
  \bibinfo{author}{\bibfnamefont{S.}~\bibnamefont{Olla}}, \bibnamefont{and}
  \bibinfo{author}{\bibfnamefont{S.~B.} \bibnamefont{Volchan}},
  \bibinfo{journal}{Commun. in Math. Phys.}
  \textbf{\bibinfo{volume}{192}}, \bibinfo{pages}{287} (\bibinfo{year}{1998}).

\bibitem[{\citenamefont{Komorowski and Olla}(2005)}]{Komorowski:2005a}
\bibinfo{author}{\bibfnamefont{T.}~\bibnamefont{Komorowski}} \bibnamefont{and}
  \bibinfo{author}{\bibfnamefont{S.}~\bibnamefont{Olla}},
  \bibinfo{journal}{J. Stat. Phys.}
  \textbf{\bibinfo{volume}{118}}, \bibinfo{pages}{407} (\bibinfo{year}{2005}).

\end{thebibliography}
\end{document}